\documentclass[preprint,showpacs,preprintnumbers,amsmath,amssymb]{revtex4}

\usepackage{graphicx}
\usepackage{dcolumn}
\usepackage{bm}

\begin{document}


\title{Where surface physics and fluid dynamics meet: rupture of an amphiphile layer by fluid flow.}

\author{M. M. Bandi}
\author{W. I. Goldburg}%
\email{goldburg@pitt.edu}
\affiliation{Department of Physics and Astronomy, University of Pittsburgh, Pittsburgh, PA 15260.}%

\author{J. R. Cressman Jr.}
\affiliation{Krasnow Institute, George Mason University, Fairfax, VA 22030.}%

\author{H. Kellay}
\affiliation{Centre de Physique Mol\'{e}culaire Optique et Hertzienne (UMR 5798),\\
Universit\'{e} Bordeaux 1, 351 cours de la Lib\'{e}ration, 33405 Talence cedex, France.}

\date{\today}

\begin{abstract}
We investigate the fluctuating pattern created by a jet of fluid impingent upon an amphiphile-covered surface. This microscopically thin layer is initially covered with 50 $\mu$m floating particles so that the layer can be visualized. A vertical jet of water located below the surface and directed upward, drives a hole in this layer. The hole is particle-free, and is surrounded by the particle-laden amphiphile region. The jet ruptures the amphiphile layer creating a particle-free region that is surrounded by the particle-covered surface. The aim of the experiment is to understand the (fluctuating) shape of the ramified interface between the particle-laden and particle-free regions.
\end{abstract}

\maketitle

\section{Introduction}
It is now well known that a monatomic layer of amphiphiles on the surface of a fluid such as water can exist in several phases \cite{dutta1999, knobler1991}. Our understanding of the properties of these phases owes so much to the work of our dear friend Chuck Knobler and his associates. This field of surface science is but one of many on which Professor Knobler has left an indelible mark.  One of us (wg)  had the good fortune to be a participant in one of these adventures.  The work presented here concerns his interest in surface physics  and issues of fluid dynamics.

This experiment concerns the behavior of particles that float on a tank of water. The basic idea can be captured by considering what happens when one stirs a spoon of powdered cream into a cup of  coffee. Assume first that the cream particles are neutrally buoyant.  Stirring will quickly disperse them through the coffee.  Once the process is complete, further stirring will have no  apparent effect.

Now  consider  a variant of that experiment in which the cream particles have a density considerably less than that of water, assuring that they will stay at the air-water interface.  Thus they experience a constraint not shared by the water molecules below. Stirring will induce the water molecules to come to the surface and to go back into the bulk, but the floaters are trapped on the surface.

Figure ~\ref{fourclouds} shows the behavior of an initially uniform distribution of floaters in a steadily stirred fluid.  Here the particles are introduced on the surface of  a large tank (lateral dimensions 1 m $\times$ 1 m) at time $t$ = 0.  Well before then, the stirring of the tank of water has been initiated, and the fluid has reached a turbulent steady state.  The motion of the particles appearing in white  Fig.  ~\ref{fourclouds},  is captured by a high speed camera looking down on the tank  from above.  The stirring is vigorous (Taylor microscale $Re_{\lambda} \simeq 150$ \cite{frisch}). One can deduce the velocity of each particle from the recorded images \cite{njp}. Because the particle distribution is uniform at $t$ = 0,   an image of the particle positions at that time would be seen as a uniformly white square and hence is not shown. The camera's field of view is limited to a square of dimensions 9.3 cm $\times$ 9.3 cm.

The images show that the surface particles are coagulating, not as spots but rather as line-like  structures.  It is seen that the coagulation is  completed in roughly 1 sec. A snapshot taken after several seconds would  be a completely black field. This is because the particles have left the region of observation and soon reach the walls of the plexi-glass tank, where they stick.

For a discussion of the technique for studying this coagulation see Ref. \cite{njp}. That reference contains details of the experiments and also simulations of the phenomenon by B. Eckhardt and J. Schumacher.  Those simulations show clearly this same coagulation phenomenon. They also demonstrate that the two dynamical Lyapunov exponents characterizing the distribution of the surface particles \cite{hilborn1994} have opposite signs.  Were they both negative, the particles would gather into point-like structures rather than string-like patterns \cite{boffetta2004prl}

Crucial to understanding this clustering effect of the floaters is a recognition that  water  is incompressible  throughout  the fluid volume, including all points, $x,y$  at the surface, $z$ = 0.  Thus $$\partial_x v_x(x,y,0,t)+ \partial_y v_y(x,y,0,t)=-\partial_z v_z(x,y,0,t) \neq 0.$$ As the floating particles exist in two dimensions, their motion is described by the left side of this equation  and hence form a compressible system. The floaters behave very differently than the water molecules on which they move.  Water molecules at the surface can acquire vertical components of velocity $v_z(x,y,0,t)$, whereas the floaters cannot.  By virtue of their buoyancy, the motion of floaters is governed by the left side of this equation, assuring for them that $\partial_x v_x(x,y,0,t) + \partial_y v_y(x,y,0,t) \neq 0$.

One may argue that because the surface is not perfectly flat, the floaters participate in three-dimensional motion by virtue of the presence of capillary waves \cite{zakharov,goldburg2001}.  This effect is small however, as was established by ancillary experiments \cite{schroder1996}.    One can define a dimensionless compressibility ${\cal C}$ that ranges from zero to unity if the turbulence is isotropic. Measurements and computer simulations establish that  ${\cal C} \simeq 0.5$ \cite{njp,boffetta2004prl}.

It is at this point in the story that surface physics enters.  The clustering that is so apparent in Fig. ~\ref{fourclouds}, is absent if  the surface is  covered with an amphiphile layer.   Therefore, the surface must be freshly ``vacuumed'' or ``skimmed'' before images like those in Fig. \ref{fourclouds} are made \cite{njp}. Typically, the ``surface vacuum cleaning'' is initiated well  before the  camera is switched on and continues through out the duration of the experiment. The skimmers are placed far from the region of observation for the surface turbulence experiments.

\section{Experiment}
The present experiments are aimed at  understanding the role of the contaminating amphiphile layer in influencing the distribution of particles on the surface. Here an amphiphile layer is deliberately placed on the surface  of a smaller tank of water (lateral dimensions 20 cm $\times$ 30 cm). That layer is ruptured by a jet of water coming from inside the tank and directed upward toward the surface (see Fig. ~\ref{small tank with jet}).  Water from the tank is fed into a small pump that supplies the vertical jet.  The orifice of the jet, located 10 cm below the surface of the water has a diameter of 1 cm.  Because the output of the  jet and its input are both inside the tank, the water level remains constant.  The flow rate from the jet is small enough so as to produce only a slight bulge ($\sim$ 2 mm) at the water surface directly above it. The diameter of the bulge is a few cm.

When the jet's vertical output hits the surface it is diverted radially outward, it's maximum speed near the center being of the order of 35 cm/s (see Fig.~\ref{radial velocity field}).  The field of view here is about 5 cm $\times$ 5 cm.   In this figure local velocities are represented by the length and direction of the small arrows. Near the center of the jet the arrows are very short, signifying that the flow is mainly upward.

It is clear that the flow is not  strongly turbulent, as it is in Fig. ~\ref{fourclouds}.  Near the edge of the image there are very few velocity vectors. This is because the surface there  is so densely covered with particles that their individual positions and velocities becomes unmeasurable.

Figure \ref{fracturedSurface} is a photo made by the overhead fast camera. The lateral dimensions of the image are about 10 cm $\times$ 10 cm. Prior to the measurement, the surface of the water is covered with a layer of  fatty acid (Oleic Acid) at close packing density of  roughly 20 $\AA^2$ per molecule.  The amphiphile layer is expected to be in a condensed phase  \cite{dutta1999, knobler1991, knobler1990}.  The white region in the center of the image is an in-plane view of the jet at the surface.  It appears to be white because a high concentration of 50 $\mu$m particles is being steadily  injected by the jet into the tank. The particles spend very little time upon reaching the surface and are quickly swept away towards the ring. The black region is therefore free of particles. Either all of the black region or an inner circle (of diameter somewhat larger than the inner white region) is likewise free of the amphiphile layer.  The particles are hollow glass spheres that show no sign of interacting with each other. Surface tension will favour coagulation of particles, but this effect is small compared to forcing effect of the jet.   

\section{Discussion}
The central problem posed by fig. \ref{fracturedSurface} is that of explaining the origin of the irregularly shaped arms or tentacles that extend into the white area.  It is believed that these particles densely cover the amphiphile layer, which is of molecular thickness and hence not visible. The average diameter of the particle-free region is about 10 cm. It is not certain if the particle free region in the dark annulus is truly amphiphile free.

A movie shows that the tentacles, as well as  the diameter of the annulus fluctuate in shape from one moment to the next, with a correlation time of the order of a fraction of a second.  If the jet is suddenly switched off, the  dark annulus collapses and the surface becomes uniformly covered with the surfactant. Presumably this tendency to uniform coverage corresponds to a lowering of the surface energy of the system. Careful observation of the collapse suggests the amphiphile layer is moving inward  so fast that the macroscopic floating  particles become detached from this microscopic layer.  As a result,radial black particle-free streaks remain, for many minutes after the (particle-free) annulus has been filled in.

So far no compelling explanation has emerged that can account for  the shape of the ramified  pattern shown in Fig.~\ref{fracturedSurface}.    The image suggests that perhaps  outward flow from the jet creates a temporally fluctuating shear that breaks the amphiphile layer, and that the floating particles sit on that  microscopically thin  layer, making it observable. If so, the flow-generated temporarily-fluctuating stress may be playing a role similar to that of the static stress in the theory of  fracture of a solid \cite{freund1998,knott1973,bonn1998,fineberg1999}.

Decades ago A. A. Griffith \cite{griffith1920} advanced a simple energy argument to explain how a fracture develops.  The mechanism is illustrated in Fig.~ \ref{fracture sketch}. Shown there is a solid, that extends into the page a distance $w$.  The solid is assumed to be homogeneous and to be characterized by the three-dimensional version of Hooke's law (stress proportional  to strain).  In the type of fracture considered here, and called a Type  I fracture,  a cut of length $a$ is made into a solid and the two exposed surfaces are pried apart with a stress $S$.

The elastic energy of the system is reduced by the lengthening of the crack but there is a surface energy cost that opposes this. A one-dimensional analogy is that of the stretching of a spring by a mass $m$ on its end. Though it costs energy $\Delta E_{1} = \frac {1}{2}ka^2$ to stretch the spring, of spring constant $k$,  there is a decrease in gravitational energy  $\Delta E_{2} = -mga$ of the mass attached to it. Since $mg = ka$, $\Delta E_{2} = -ka^2$. Therefore the total energy change is $\Delta E_{b} = \Delta E_{1} + \Delta E_{2} = -\frac{1}{2}ka^2$, a negative quantity.  In both 1D and 2D, the energy {\underline {decrease}} is proportional to $a^2$ if the solid obeys Hooke's law. However, as the crack grows, there is a positive contribution $2 \gamma  a w $, where $\gamma$ is a surface energy coefficient, or surface tension  (the crack has two faces, hence the factor of two).  Figure~\ref{griffplot} shows the energy contributions of the two terms in the Griffith theory.  Beyond a critical length $a_c$, a crack will spontanteously grow; for smaller lengths it will heal. At  $a=a_c$ the system is in a state of unstable equilibrium.

For brittle fracture of a 3D solid of Young's modulus $E$, the stress $\sigma_c$ required to make a crack of length $a$ marginally unstable to spontaneous growth, is given by $$\sigma= A  \sqrt {E \gamma /a},$$ where $A$ is a constant  of order  unity  \cite{fineberg1999}, $E$ is Young's modulus (in Pa), and $\gamma$ is the surface energy (in N/m) required to create a fracture.   One may ask if this expression can be applied to the fracture of a 2D amhiphile layer, with $E$ and $\gamma$  replaced by their two-dimensional counterparts.   Both the  2D Young's modulus  $E_2$ of an amphiphile  layer and the corresponding surface energy $\gamma_2$ have been measured, though not in oleic acid.   Roughly speaking these coefficients are in the range $E_2 \simeq$ 20 mN/m  (in  pentadodecanoic acid \cite{hatta2002}) and $\gamma_2  \simeq $ 30 pN (in methyl-octadecanoate  \cite{wurlitzer2000}).  The fingers in fig. \ref{fracturedSurface} have a length of the order of 1 cm. Inserting this parameter in the above equation and using the above values of $E_2$ and $\gamma_2$ one estimates a critical stress of the amphiphile layer $\sigma_c$ of the order of $10^{-5}$ N/m.

It is assumed that the stress that ruptures the amphiphile layer originates from the radial flow. The source of the corresponding two-dimensional stress, now called $\sigma_2$, is the viscous shear on the amphiphile covering.  That stress is taken to be $\sigma_2 =(2 \pi R) \eta \frac{\Delta U}{\Delta y}$, where $\Delta U$ is taken to be the difference between the radial velocity of the amphiphile layer and the flow rate near the surface when the amphiphile layer is absent.  Its value is roughly $\Delta U$ = 10 cm/s. This gradient is across a vertical distance $\Delta y$ that has not been measured; it is estimated to be $\Delta y$ = 1 cm (the jet originates a distance of 5  cm below the surface). Taking $\eta = 0.01$ poise for water, and  radius of the amphiphile-free hole is $R=5 cm$ roughly, thus giving a crude estimate of the two-dimensional stress $\sigma_2 \simeq 10^{-3}$ N/m, two orders of magnitude larger than $\sigma_c = 10^{-5}$ N/m.  If these estimates are to be trusted, the radial shear on the amphiphile layer is more than sufficient to create fissures into the amphiphile-covered (and
particle-covered) region in Fig. 4.

Though the fjord-like channels in Fig. ~\ref{fracturedSurface} do not resemble fracture lines seen in solids,  it has been found  that a hydrophobic layer of  lycopodium powder is fractured by an amphiphile  introduced on it \cite{vella2004}. It has also been suggested that the phenomenon observed in this experiment is related to the viscous fingering instability \cite{saffmantaylor}. It appears however that this explanation requires that the underlying fluid be microscopically shallow (private communication with S. M. Troian and \cite{troian1990}), which seems to rule out viscous fingering. It is suggested above that the ramified structure observed in figure \ref{fracturedSurface} is related to a balance between the stress applied by the jet  and the  one-dimensional surface energy of the amphiphile layer. The irregular structure of the intrusions may arise because the flow is chaotic rather than laminar. It must be noted that unlike the fracture in brittle solids, the tentacles observed in this experiment are initially very blunt and only later evolve into sharper tips.
 
\section{Summary}
Particles floating on the freshly cleaned  surface of a tank of turbulent water coagulate in a way that is expected for a compressible fluid. However, if that surface is  covered with an amphiphile layer, this coagulation effect is  blocked.  The experiments described here are aimed  at understanding the role of  this layer in a fluid dynamics setting. That layer is
rendered visible by a covering of  small particles that float on the water's surface. It comes as no surprise that the jet will break a hole in the particle covering.  If the surface is amphiphile-free, that hole is smooth and expands to the edge of the tank.
On the other hand, if the amphiphile covering is present, this hole develops irregularly shaped tendrils at its rim, and these  particle-free regions fluctuate in space and time.  An admittedly crude argument suggests that the observed phenomenon may be understood in terms of the theory of fracture.  Only further and more extensive experiments will determine if the fracture approach has merit.  In any case, the observed patterns are intriguing and call for an explanation.

\begin{acknowledgments}
This work was supported by the National Science Foundation grant DMR No. 0201805. The authors have benefited greatly from discussions with C. M. Knobler, Thomas Fischer, Sandra Troian and the University of Pittsburgh Softmatter group. The authors also acknowledge Michael Turner for help with initial measurements.
\end{acknowledgments}



\bibliography{monolayer01112005.bib}

\begin{thebibliography}{20}
\expandafter\ifx\csname natexlab\endcsname\relax\def\natexlab#1{#1}\fi
\expandafter\ifx\csname bibnamefont\endcsname\relax
  \def\bibnamefont#1{#1}\fi
\expandafter\ifx\csname bibfnamefont\endcsname\relax
  \def\bibfnamefont#1{#1}\fi
\expandafter\ifx\csname citenamefont\endcsname\relax
  \def\citenamefont#1{#1}\fi
\expandafter\ifx\csname url\endcsname\relax
  \def\url#1{\texttt{#1}}\fi
\expandafter\ifx\csname urlprefix\endcsname\relax\def\urlprefix{URL }\fi
\providecommand{\bibinfo}[2]{#2}
\providecommand{\eprint}[2][]{\url{#2}}

\bibitem[{\citenamefont{Kaganer and Dutta}(1999)}]{dutta1999}
\bibinfo{author}{\bibfnamefont{V.~M.} \bibnamefont{Kaganer}} \bibnamefont{and}
  \bibinfo{author}{\bibfnamefont{P.}~\bibnamefont{Dutta}},
  \bibinfo{journal}{Rev. Mod. Phys.} \textbf{\bibinfo{volume}{71}},
  \bibinfo{pages}{779} (\bibinfo{year}{1999}).

\bibitem[{\citenamefont{Knobler}(1991)}]{knobler1991}
\bibinfo{author}{\bibfnamefont{C.~M.} \bibnamefont{Knobler}},
  \bibinfo{journal}{J. Phys: Condens. Matter} \textbf{\bibinfo{volume}{3}},
  \bibinfo{pages}{S17} (\bibinfo{year}{1991}).

\bibitem[{\citenamefont{Frisch}(1995)}]{frisch}
\bibinfo{author}{\bibfnamefont{U.}~\bibnamefont{Frisch}},
  \bibinfo{journal}{Cambridge University Press, Cambridge}
  (\bibinfo{year}{1995}).

\bibitem[{\citenamefont{Cressman et~al.}(2004)\citenamefont{Cressman, Davoudi,
  Goldburg, and Schumacher}}]{njp}
\bibinfo{author}{\bibfnamefont{J.~R.} \bibnamefont{Cressman}},
  \bibinfo{author}{\bibfnamefont{J.}~\bibnamefont{Davoudi}},
  \bibinfo{author}{\bibfnamefont{W.~I.} \bibnamefont{Goldburg}},
  \bibnamefont{and}
  \bibinfo{author}{\bibfnamefont{J.}~\bibnamefont{Schumacher}},
  \bibinfo{journal}{New Journal of Physics} \textbf{\bibinfo{volume}{6}},
  \bibinfo{pages}{53} (\bibinfo{year}{2004}).

\bibitem[{\citenamefont{Hilborn}(1994)}]{hilborn1994}
\bibinfo{author}{\bibfnamefont{R.~C.} \bibnamefont{Hilborn}},
  \bibinfo{journal}{Oxford U. Press, New York} p.~\bibinfo{pages}{1}
  (\bibinfo{year}{1994}).

\bibitem[{\citenamefont{Boffetta et~al.}(2004)\citenamefont{Boffetta, Davoudi,
  Eckhardt, and Schumacher}}]{boffetta2004prl}
\bibinfo{author}{\bibfnamefont{G.}~\bibnamefont{Boffetta}},
  \bibinfo{author}{\bibfnamefont{J.}~\bibnamefont{Davoudi}},
  \bibinfo{author}{\bibfnamefont{B.}~\bibnamefont{Eckhardt}}, \bibnamefont{and}
  \bibinfo{author}{\bibfnamefont{J.}~\bibnamefont{Schumacher}},
  \bibinfo{journal}{Phys. Rev. Lett.} \textbf{\bibinfo{volume}{93}},
  \bibinfo{pages}{134501} (\bibinfo{year}{2004}).

\bibitem[{\citenamefont{Zakharov et~al.}(1992)\citenamefont{Zakharov, L'vov,
  and Falkovich}}]{zakharov}
\bibinfo{author}{\bibfnamefont{V.~E.} \bibnamefont{Zakharov}},
  \bibinfo{author}{\bibfnamefont{V.}~\bibnamefont{L'vov}}, \bibnamefont{and}
  \bibinfo{author}{\bibfnamefont{G.}~\bibnamefont{Falkovich}},
  \bibinfo{journal}{Springer Series in NONLINEAR DYNAMICS, Springer-Verlag (New
  York)}  (\bibinfo{year}{1992}).

\bibitem[{\citenamefont{Goldburg et~al.}(2001)\citenamefont{Goldburg, Cressman,
  V\"or\"os, Eckhardt, and Schumacher}}]{goldburg2001}
\bibinfo{author}{\bibfnamefont{W.~I.} \bibnamefont{Goldburg}},
  \bibinfo{author}{\bibfnamefont{J.~R.} \bibnamefont{Cressman}},
  \bibinfo{author}{\bibfnamefont{Z.}~\bibnamefont{V\"or\"os}},
  \bibinfo{author}{\bibfnamefont{B.}~\bibnamefont{Eckhardt}}, \bibnamefont{and}
  \bibinfo{author}{\bibfnamefont{J.}~\bibnamefont{Schumacher}},
  \bibinfo{journal}{Phys. Rev. E} \textbf{\bibinfo{volume}{63}},
  \bibinfo{pages}{065303} (\bibinfo{year}{2001}).

\bibitem[{\citenamefont{Schr\"{o}der et~al.}(1996)\citenamefont{Schr\"{o}der,
  Andersen, Levinsen, Alstr\"{o}m, and Goldburg}}]{schroder1996}
\bibinfo{author}{\bibfnamefont{E.}~\bibnamefont{Schr\"{o}der}},
  \bibinfo{author}{\bibfnamefont{J.~S.} \bibnamefont{Andersen}},
  \bibinfo{author}{\bibfnamefont{M.~T.} \bibnamefont{Levinsen}},
  \bibinfo{author}{\bibfnamefont{P.}~\bibnamefont{Alstr\"{o}m}},
  \bibnamefont{and} \bibinfo{author}{\bibfnamefont{W.~I.}
  \bibnamefont{Goldburg}}, \bibinfo{journal}{Phys. Rev. Lett.}
  \textbf{\bibinfo{volume}{76}}, \bibinfo{pages}{4717} (\bibinfo{year}{1996}).

\bibitem[{\citenamefont{Knobler}(1990)}]{knobler1990}
\bibinfo{author}{\bibfnamefont{C.~M.} \bibnamefont{Knobler}},
  \bibinfo{journal}{Adv. Chem. Phys.} \textbf{\bibinfo{volume}{77}},
  \bibinfo{pages}{397} (\bibinfo{year}{1990}).

\bibitem[{\citenamefont{Freund}(1998)}]{freund1998}
\bibinfo{author}{\bibfnamefont{L.~B.} \bibnamefont{Freund}},
  \bibinfo{journal}{Cambridge Univ. Press, Cambridge}  (\bibinfo{year}{1998}).

\bibitem[{\citenamefont{Knott}(1973)}]{knott1973}
\bibinfo{author}{\bibfnamefont{J.~F.} \bibnamefont{Knott}},
  \bibinfo{journal}{Butterworth}  (\bibinfo{year}{1973}).

\bibitem[{\citenamefont{Bonn et~al.}(1998)\citenamefont{Bonn, Kellay, Prochnow,
  Ben-Djemiaa, and Meunier}}]{bonn1998}
\bibinfo{author}{\bibfnamefont{D.}~\bibnamefont{Bonn}},
  \bibinfo{author}{\bibfnamefont{H.}~\bibnamefont{Kellay}},
  \bibinfo{author}{\bibfnamefont{M.}~\bibnamefont{Prochnow}},
  \bibinfo{author}{\bibfnamefont{K.}~\bibnamefont{Ben-Djemiaa}},
  \bibnamefont{and} \bibinfo{author}{\bibfnamefont{J.}~\bibnamefont{Meunier}},
  \bibinfo{journal}{Science} \textbf{\bibinfo{volume}{280}},
  \bibinfo{pages}{265} (\bibinfo{year}{1998}).

\bibitem[{\citenamefont{Fineberg and Marder}(1999)}]{fineberg1999}
\bibinfo{author}{\bibfnamefont{J.}~\bibnamefont{Fineberg}} \bibnamefont{and}
  \bibinfo{author}{\bibfnamefont{M.}~\bibnamefont{Marder}},
  \bibinfo{journal}{Phys. Rept.} \textbf{\bibinfo{volume}{313}},
  \bibinfo{pages}{1} (\bibinfo{year}{1999}).

\bibitem[{\citenamefont{Griffith}(1920)}]{griffith1920}
\bibinfo{author}{\bibfnamefont{A.~A.} \bibnamefont{Griffith}},
  \bibinfo{journal}{Phil. Trans. Roy. Soc., London}
  \textbf{\bibinfo{volume}{Ser. A221}}, \bibinfo{pages}{163}
  (\bibinfo{year}{1920}).

\bibitem[{\citenamefont{Hatta and Fischer}(2002)}]{hatta2002}
\bibinfo{author}{\bibfnamefont{E.}~\bibnamefont{Hatta}} \bibnamefont{and}
  \bibinfo{author}{\bibfnamefont{T.~M.} \bibnamefont{Fischer}},
  \bibinfo{journal}{Langmiur} \textbf{\bibinfo{volume}{18}},
  \bibinfo{pages}{6201} (\bibinfo{year}{2002}).

\bibitem[{\citenamefont{Wurlitzer et~al.}(2000)\citenamefont{Wurlitzer,
  Steffen, Wurlitzer, Khattari, and Fischer}}]{wurlitzer2000}
\bibinfo{author}{\bibfnamefont{S.}~\bibnamefont{Wurlitzer}},
  \bibinfo{author}{\bibfnamefont{P.}~\bibnamefont{Steffen}},
  \bibinfo{author}{\bibfnamefont{M.}~\bibnamefont{Wurlitzer}},
  \bibinfo{author}{\bibfnamefont{Z.}~\bibnamefont{Khattari}}, \bibnamefont{and}
  \bibinfo{author}{\bibfnamefont{T.~M.} \bibnamefont{Fischer}},
  \bibinfo{journal}{J. Chem. Phys.} \textbf{\bibinfo{volume}{113}},
  \bibinfo{pages}{3822} (\bibinfo{year}{2000}).

\bibitem[{\citenamefont{Vella et~al.}(2004)\citenamefont{Vella, Aussillours,
  and Mahadevan}}]{vella2004}
\bibinfo{author}{\bibfnamefont{D.}~\bibnamefont{Vella}},
  \bibinfo{author}{\bibfnamefont{P.}~\bibnamefont{Aussillours}},
  \bibnamefont{and}
  \bibinfo{author}{\bibfnamefont{L.}~\bibnamefont{Mahadevan}},
  \bibinfo{journal}{Europhys Lett} \textbf{\bibinfo{volume}{68}},
  \bibinfo{pages}{212} (\bibinfo{year}{2004}).

\bibitem[{\citenamefont{Saffman and Taylor}(1958)}]{saffmantaylor}
\bibinfo{author}{\bibfnamefont{P.~G.} \bibnamefont{Saffman}} \bibnamefont{and}
  \bibinfo{author}{\bibfnamefont{G.~I.} \bibnamefont{Taylor}},
  \bibinfo{journal}{Proc. Roy. Soc. Lond. A} \textbf{\bibinfo{volume}{245}},
  \bibinfo{pages}{312} (\bibinfo{year}{1958}).

\bibitem[{\citenamefont{Troian et~al.}(1990)\citenamefont{Troian,
  Herbolzheimer, and Safran}}]{troian1990}
\bibinfo{author}{\bibfnamefont{S.~M.} \bibnamefont{Troian}},
  \bibinfo{author}{\bibfnamefont{E.}~\bibnamefont{Herbolzheimer}},
  \bibnamefont{and} \bibinfo{author}{\bibfnamefont{S.~A.}
  \bibnamefont{Safran}}, \bibinfo{journal}{Phys. Rev. Lett.}
  \textbf{\bibinfo{volume}{65}}, \bibinfo{pages}{333} (\bibinfo{year}{1990}).

\end{thebibliography}

\clearpage

\begin{figure}
\begin{center}
\includegraphics[width=4in]{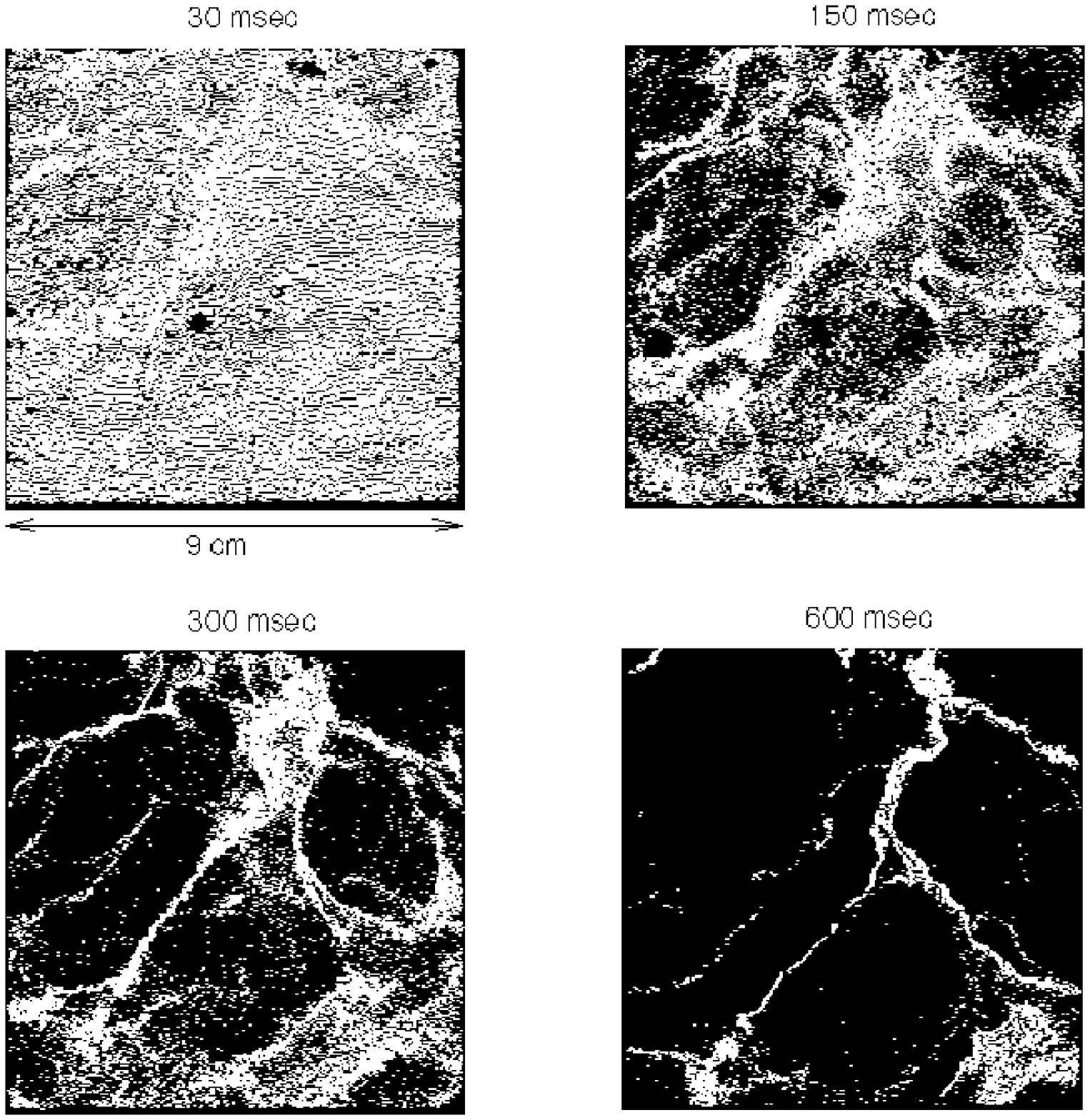}
\end{center}
\caption{\label{fourclouds}}
\end{figure}

\begin{figure}
\begin{center}
\includegraphics[width = 4 in]{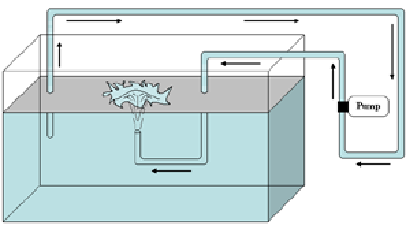}
\end{center}
\caption{\label{small tank with jet}}
\end{figure}

\begin{figure}
\begin{center}
\includegraphics[width = 4 in]{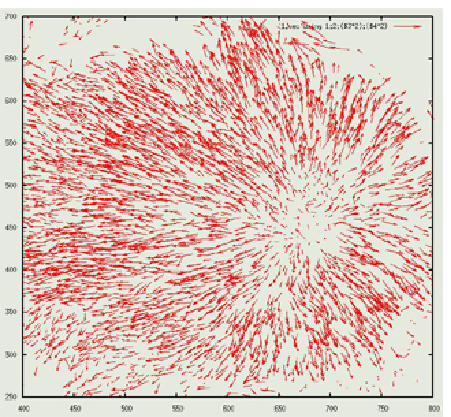}
\end{center}
\caption{\label{radial velocity field}}
\end{figure}

\begin{figure} 
\begin{center} 
\includegraphics[width = 4 in]{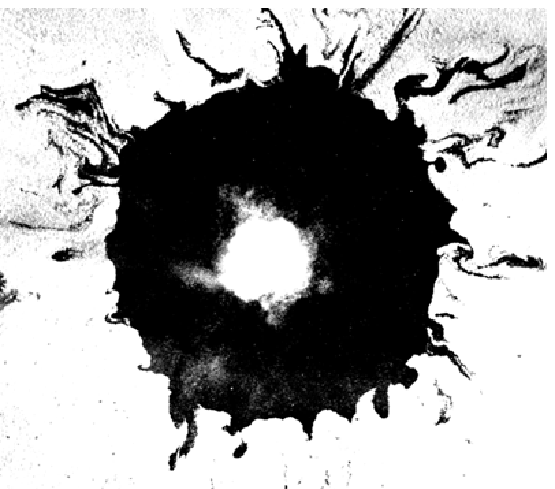} 
\end{center} \caption{\label{fracturedSurface}}
\end{figure}

\begin{figure}
\begin{center}
\includegraphics[width = 4 in]{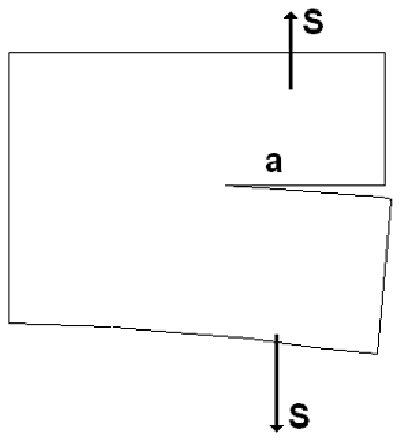}
\end{center}
\caption{\label{fracture sketch}}
\end{figure}

\begin{figure}
\begin{center}
\includegraphics[width = 5 in]{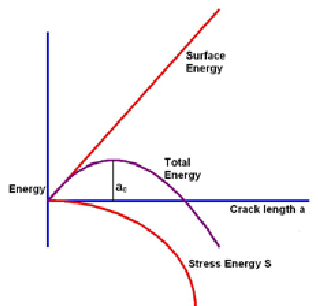}
\end{center}
\caption{\label{griffplot}}
\end{figure}

{\renewcommand
\baselinestretch 2
Fig \ref{fourclouds}: Evolution of particle density of floaters that are initially distributed uniformly at $t$=0.  Tracers appear in white.  For these measurements, $Re_{\lambda }$ = 140, $\lambda$ = 0.4 cm, integral  scale, $l_0$ = 3.8 cm.
\newline

Fig \ref{small tank with jet}:  A sketch of the experimental setup. The tank has lateral dimensions of 20 cm $\times$ 30 cm and is filled with water to a height of about 30 cm. The jet is placed 10 cm below the surface. An amphiphile monolayer is introduced on the surface prior to the experiment. A micropump circulates water from the tank through the jet rupturing the monolayer. Particles constantly seeded onto the surface from the jet allow visualization of the ruptured interface between the water and amphiphilic monolayer.
\newline

Fig \ref{radial velocity field}: Velocity vector field of the radial flow at the surface as measured by Particle Imaging Velocimetry \cite{njp}. The field of view is 5 cm $\times$ 5 cm.
\newline

Fig \ref{fracturedSurface}: A snapshot showing ramification of the amphiphilic monolayer at the air-water interface. In the steady state this pattern fluctuates in time, but the mean diameter of the dark annulus remains roughly constant. The tentacles change in depth, orientation and shape. Their typical length is of the order of $a = 1$ cm.
\newline

Fig \ref{fracture sketch}: Type 1 fracture of a three-dimensional solid. For a given stress $S$, the crack length $a$ will spontaneously grow if $S$ exceeds a critical value $S_{c}$, where $S_{c}$ depends upon the Young's modulus of the solid and the surface energy increase that comes from creating the crack. If $S$ is set to a particular value,a crack of length $a$ will spontaneously grow only if $a$ exceeds some critical value $a=a_c$. If $a < a_c$, the crack heals.
\newline

Fig \ref{griffplot}: Type 1 fracture of a three-dimensional solid. For a given stress $S$, the crack length $a$ will spontaneously grow if $S$ exceeds a critical value $S_{c}$, where $S_{c}$ depends upon the Young's modulus of the solid and the surface energy increase that comes from creating the crack. If $S$ is set to a particular value,a crack of length $a$ will spontaneously grow only if $a$ exceeds some critical value $a=a_c$. If $a < a_c$, the crack heals.
}
\clearpage

\clearpage

\clearpage

\clearpage

\clearpage

\clearpage

\end{document}